\documentclass[authoryear,preprint,review,12pt]{elsarticle}

\usepackage{geometry}
\usepackage{fancyhdr}
\usepackage{titlesec}
\usepackage{float}
\usepackage{subcaption}

\usepackage{amsmath,amsfonts,amssymb}
\usepackage{mathtools}
\usepackage{bm}
\usepackage{mathrsfs}

\usepackage{graphicx}
\usepackage{xcolor}
\usepackage{tikz}
\usepackage{pgfplots}

\usepackage{tabularx}
\usepackage{multirow}
\usepackage{longtable}
\usepackage{colortbl}
\usepackage{booktabs}

\usepackage{enumitem}

\usepackage{hyperref}
\usepackage{cleveref}

\usepackage{natbib}

\usepackage{microtype}
\usepackage{siunitx}
\usepackage{caption}
\usepackage{subcaption}
\usepackage{lipsum}
\usepackage{xspace}

\journal{Physics Open}
\pdfoutput=1
\pgfplotsset{compat=1.18}
\begin{document}

\begin{frontmatter}

%% Title, authors and addresses

%% use the tnoteref command within \title for footnotes;
%% use the tnotetext command for theassociated footnote;
%% use the fnref command within \author or \affiliation for footnotes;
%% use the fntext command for theassociated footnote;
%% use the corref command within \author for corresponding author footnotes;
%% use the cortext command for theassociated footnote;
%% use the ead command for the email address,
%% and the form \ead[url] for the home page:
%% \title{Title\tnoteref{label1}}
%% \tnotetext[label1]{}
%% \author{Name\corref{cor1}\fnref{label2}}
%% \ead{email address}
%% \ead[url]{home page}
%% \fntext[label2]{}
%% \cortext[cor1]{}
%% \affiliation{organization={},
%%            addressline={}, 
%%            city={},
%%            postcode={}, 
%%            state={},
%%            country={}}
%% \fntext[label3]{}
\title{Uncertainty Quantification of Bandgaps in Acoustic Metamaterials with Stochastic Geometric Defects and Material Properties}
%\title{Uncertainty Quantification of Acoustic Dispersion Bands in Metamaterials with Stochastic Geometric Defects and Material Properties}
%\title{Uncertainty Quantification of Acoustic Dispersion Bands in Metamaterials}

%\title{Quantifying Uncertainty in Metamaterial Dispersion Bands With Stochastic Geometric Defects and Material Properties}
%\title{Quantifying Uncertainty in Metamaterial Dispersion Bands}

%% use optional labels to link authors explicitly to addresses:
%% \author[label1,label2]{}
%% \affiliation[label1]{organization={},
%%             addressline={},
%%             city={},
%%             postcode={},
%%             state={},
%%             country={}}
%%
%% \affiliation[label2]{organization={},
%%             addressline={},
%%             city={},
%%             postcode={},
%%             state={},
%%             country={}}

\author[duke]{Han Zhang}
\author[duke]{Rayehe Karimi Mahabadi}
\author[duke]{Cynthia Rudin}
\author[duke]{Johann Guilleminot}
\author[duke]{L. Catherine Brinson}
\affiliation[duke]{organization={Duke University},%Department and Organization
            %addressline={}, 
            city={Durham},
            %postcode={}, 
            state={NC},
            country={USA}}
\begin{abstract}
%% Text of abstract
This paper studies the utility of techniques within uncertainty quantification, namely spectral projection and polynomial chaos expansion, in reducing sampling needs for characterizing acoustic metamaterial dispersion band responses given stochastic material properties and geometric defects. A novel method of encoding geometric defects in an interpretable, resolution independent is showcased in the formation of input space probability distributions. Orders of magnitude sampling reductions down to $\sim10^0$ and $\sim10^1$ are achieved in the 1D and 7D input space scenarios respectively while maintaining accurate output space probability distributions through combining Monte Carlo, quadrature rule, and sparse grid sampling with surrogate model fitting.
%\todo{lengthen, by including more of method and conclusion}
\end{abstract}

%%Graphical abstract
%\begin{graphicalabstract}
%\includegraphics{grabs}
%\end{graphicalabstract}

%%Research highlights
%\begin{highlights}
%\item Research highlight 1
%\item Research highlight 2
%\end{highlights}

\begin{keyword}
%% keywords here, in the form: keyword \sep keyword, up to a maximum of 6 keywords
1\sep Uncertainty Quantification 2\sep Metamaterials, 3\sep Acoustics, 4\sep spectral projection, 5\sep Polynomial Chaos Expansion
%% PACS codes here, in the form: \PACS code \sep code

%% MSC codes here, in the form: \MSC code \sep code
%% or \MSC[2008] code \sep code (2000 is the default)

\end{keyword}

\end{frontmatter}

%\tableofcontents

%% \linenumbers

%% main text
%%%%%%%%%%%%%%%%%%%%%%%%%%%%%%%%%%%%%%%%%%%%%%%%%%%%%%%%%%%%%%%%%%%%%%%%%%%%%%%%%%%%%%%%%%%%%%%%%%%%%%%%%%%%%%%%%%%%%
\section{Introduction}
\label{introduction}
Acoustic metamaterials have gained enormous attention in recent decades for their capabilities in manipulating vibration waves in myriad ways as a consequence of their design geometry and material properties \citep{review2020}. These theoretically arbitrary wave manipulations understandably have much applicability in many areas of engineering and science, as well as commercial use. For all these real world applications, there is intrinsic variability in material properties and geometry deviations from designed parameters, which affect the acoustic properties of the metamaterial. Therefore, the question of how to quantify the effects of stochastic material properties and geometry defects is a natural one to raise. In this paper, we will see the applicability of methods like spectral projection and polynomial chaos expansion in capturing the variability propagating from stochastic inputs like material properties and a novel resolution invariant geometry defect parameter, through a ground truth model, to the output wave dispersion characteristics. These methods may be of use to the reader in reducing their sampling needs to reach an arbitrary level of confidence in their metamaterial performance characteristics.

\section{Methodology}
%%%%%%%%%%%%%%%%%%%%%%%%%%%%%%%%%%%%%%%%%%%%%%%%%%%%%%%%%%%%%%%%%%%%%%%%%%%%%%%%%%%%%%%%%%%%%%%%%%%%%%%%%%%%%%%%%%%%%
\subsection{Acoustic Metamaterials}
Here we provide a brief overview of how acoustic metamaterials work, and how we compute the effects of material properties and geometry of given metamaterials. 
%%%%%%%%%%%%%%%%%%%%%%%%%%%%%%%%%%%%%%%%%%%%%%%%%%%%%%%%%%%%%%%%%%%%%%%%%%%%%%%%%%%%%%%%%%%%%%%%%%%%%%%%%%%%%%%%%%%%%
\subsubsection{Material Properties}
By definition, an acoustic metamaterial achieves its acoustic function through some combination of its geometry and its material properties. These material properties influence the propagation of vibration waves through the material, which will be detailed in section \cref{subsubsection:Dispersion Relations and Bandgaps}. The metamaterials in the following studies are all comprised of two materials, with properties shown below in \cref{table:Material Properties}. Note that the stiffer of the two materials is representative of a steel alloy, and the softer of the two materials is representative of a cured epoxy resin. These two sets of material properties were chosen because of a known bandgap presence from previous work by \citep{CHEN2022101895} which makes it expedient to set up a uncertainty quantification problem, and finally because this combination is readily manufacturable should the need for real world validation ever arise.
\begin{table}[H]
\centering
\small
%\tiny
% \caption{7D Gamma Material Property \& Beta Geometry Input Distributions}
\caption{7D Material Property \& Geometry Input Distributions}
\label{table:Material Properties}
\begin{tabularx}{\linewidth}{Xcc} 
\toprule
Material Property & Soft Material Nominal Value & Hard Material Nominal Value \\
\midrule
\( \text{Bulk Modulus } (K)\) & \( 278 \text{ MPa} \) & \( 152 \text{ GPa} \) \\
\( \text{Shear Modulus } (G)\) & \( 72.5 \text{ MPa} \) & \( 78.1 \text{ GPa} \) \\
\( \text{Young's Modulus } (E)\) & \( 200 \text{ MPa} \) & \( 200 \text{ GPa} \) \\
\( \text{Poisson Ratio } (\nu)\) & \( 0.38 \) & \( 0.28 \) \\
\( \text{Density } (\rho)\) & \( 1000 \text{ g/cm$^3$} \) & \( 8000 \text{ g/cm$^3$} \) \\
\bottomrule
\end{tabularx}
\end{table}
%%%%%%%%%%%%%%%%%%%%%%%%%%%%%%%%%%%%%%%%%%%%%%%%%%%%%%%%%%%%%%%%%%%%%%%%%%%%%%%%%%%%%%%%%%%%%%%%%%%%%%%%%%%%%%%%%%%%%
\subsubsection{Geometry}
The metamaterials in this paper are 2D geometries, which are represented as a matrix of 0s and 1s, representing placements of soft and hard material respectively. These matrices can be visualized like in \cref{fig:design_geos} below, with black pixels representing 1s, and white pixels representing 0s.
\begin{figure}[H]
    \centering
    \begin{subfigure}[b]{0.49\linewidth}
        \centering
        \includegraphics[width=\linewidth]{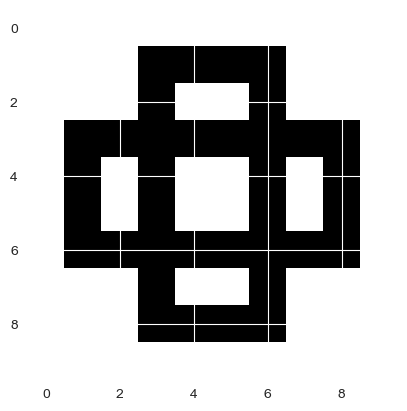}
    \end{subfigure}
    \hfill
    \begin{subfigure}[b]{0.49\linewidth}
        \centering
        \includegraphics[width=\linewidth]{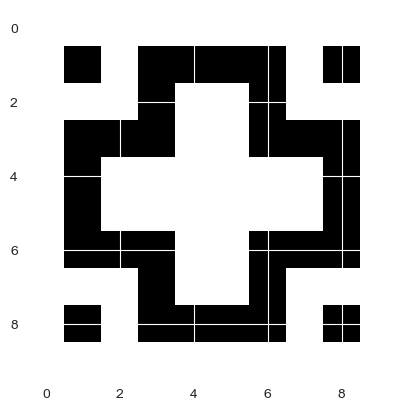}
    \end{subfigure}
    \caption{The two design geometries used in the following studies.}
    \label{fig:design_geos}    
\end{figure}
%%%%%%%%%%%%%%%%%%%%%%%%%%%%%%%%%%%%%%%%%%%%%%%%%%%%%%%%%%%%%%%%%%%%%%%%%%%%%%%%%%%%%%%%%%%%%%%%%%%%%%%%%%%%%%%%%%%%%
\subsubsection{Dispersion Relations and Bandgaps}
\label{subsubsection:Dispersion Relations and Bandgaps}
Given the material properties and geometry of an acoustic metamaterial, its dispersion relation can be calculated by solving the Navier equations \citep{zhang2021realization}, presented as follows 
\begin{equation}
    (\lambda (\mathbf{r})+ G (\mathbf{r})) \nabla(\nabla \cdot \mathbf{u}) + G (\mathbf{r}) \nabla^2 \mathbf{u} = \rho (\mathbf{r}) \frac{\partial^2 \mathbf{u}}{\partial t^2} \label{eq1}
\end{equation} 
where $\lambda$ is the Lamé constant, $G$ is the shear modulus, $\mathbf{u}$ is the displacement vector, and $\rho$ is the density. 
Because regular metamaterials are formed by tiling a unit cell, we can reduce the domain of our analysis in \cref{eq1} to a unit cell under Bloch-Floquet periodic boundary conditions. According to Bloch-Floquet theory, in a periodic domain, we can express the displacement field as \citep{yang2016effective}
\begin{equation}
    \mathbf{u}(\mathbf{r}+\mathbf{a}) = \mathbf{u}(\mathbf{r})e^{i\mathbf{k} \cdot \mathbf{a}}  \label{eq2}
\end{equation}
where $\mathbf{r}$ is the position vector, $\mathbf{a}$ is the lattice vector, and $\mathbf{k}$ is the wave vector. To solve \cref{eq1} in our reduced domain, we discretize a unit cell using the finite element method and rewrite the formulation as a generalized eigenvalue problem.
\begin{equation}
    (\mathbf{K}(\mathbf{k}) - \omega^2 \mathbf{M})\mathbf{U} = \mathbf{0}  \label{eq3}
\end{equation}
where $\mathbf{K}$ and $\mathbf{M}$ are the stiffness and mass matrices of the unit cell respectively. By solving \cref{eq3} with a finite element analysis (FEA) solver, we can find the relation between eigenfrequency ($\omega$) and wave vector ($\mathbf{k}$), which forms the dispersion curve. By examining the dispersion curves, we can identify bandgaps where propagation of waves with certain frequencies is prohibited by the metamaterial. These bandgaps will be characterized in the following studies by their center frequency and bandwidth.

%%%%%%%%%%%%%%%%%%%%%%%%%%%%%%%%%%%%%%%%%%%%%%%%%%%%%%%%%%%%%%%%%%%%%%%%%%%%%%%%%%%%%%%%%%%%%%%%%%%%%%%%%%%%%%%%%%%%%
\subsection{Uncertainty Quantification}
The purpose of uncertainty quantification (UQ) is to endow predictions with some probabilistic measure of confidence \citep{ghanem2017handbook}. One important aspect in UQ concerns the propagation of uncertainties where the impact of stochastic inputs on quantities of interest is quantified (given some input-output model $f$). 
% , given some model and stochastic input space to that model, predict how the input stochasticity will propagate to the output space. In other words, UQ aims to map an input probability density to an output probability density. 
Let $\mathbf{X} = (X_1, X_2, \dots, X_m)^T$ represent our stochastic $m$-dimensional input (with statistically independent components), defined on a probability space $(\Theta, \mathcal{F}, P)$. We denote by $P_{\mathbf{X}}$ the probability measure of $\mathbf{X}$, defined by the probability density function $p_{\mathbf{X}}$ with respect to the Lebesgue measure $d\mathbf{x}$ in $\mathbb{R}^m$: $P_{\mathbf{X}}(d\mathbf{x}) = p_{\mathbf{X}}(\mathbf{x})d\mathbf{x}$. Let $P_{\mathbf{Y}}$ be the pushed-forward (i.e., image) measure through $f$, associated with the stochastic $q$-dimensional output $\mathbf{Y} = f(\mathbf{X})$. We denote by $p_{\mathbf{Y}}$ the probability density function defining $P_{\mathbf{Y}}$.
% Suppose also that our model $f$ produces outputs $\mathbf{Y} = f(\mathbf{X})$ with an inherited probability density function $P(\mathbf{y})$ due to the stochasticity in $\mathbf{X}$. 
The task is then to estimate $P_{\mathbf{Y}}$, given $P_{\mathbf{X}}$. This can be achieved through various techniques, including Monte Carlo sampling and surrogate modeling methods. %In the studies that follow, we will demonstrate the utility of such approaches in capturing the stochasticity of acoustic metamaterial bandgap locations and sizes, given stochastic material properties, and geometric defects of the metamaterial unit cell.
%%%%%%%%%%%%%%%%%%%%%%%%%%%%%%%%%%%%%%%%%%%%%%%%%%%%%%%%%%%%%%%%%%%%%%%%%%%%%%%%%%%%%%%%%%%%%%%%%%%%%%%%%%%%%%%%%%%%%
\subsubsection{Monte Carlo Sampling}
The Monte Carlo (MC) approach involves the following steps:
\begin{enumerate}
    \item Generate a large number N of random samples $\mathbf{X}_1, \mathbf{X}_2, \dots, \mathbf{X}_N$, drawn from $P_{\mathbf{X}}$.
    \item Evaluate the associated output samples: $\mathbf{Y}_i = f(\mathbf{X}_i) \quad \text{for } i = 1,2, \dots, N$.
    \item Estimate $p_{\mathbf{Y}}$ (using a kernel density estimate, for instance) and/or analyze statistical moments of $\mathbf{Y}$ (e.g., the mean and the covariance matrix).
\end{enumerate}

% Monte Carlo sampling is always guaranteed to give an accurate reflection of the true probability distribution with a large enough sample size but is usually the least efficient in terms of samples required and therefore computation costs. 
Whenever applicable (due to the computational associated with model evaluation for random instances of $\mathbf{X}$), this approach delivers baseline results that can be used to assess the relevance of alternative techniques, including the ones demonstrated in this paper.
%%%%%%%%%%%%%%%%%%%%%%%%%%%%%%%%%%%%%%%%%%%%%%%%%%%%%%%%%%%%%%%%%%%%%%%%%%%%%%%%%%%%%%%%%%%%%%%%%%%%%%%%%%%%%%%%%%%%%
\subsubsection{Polynomial Chaos Expansion (PCE)}
Assume that $\mathbf{Y} = f(\mathbf{X})$ is a second-order random variable, with $E\{\|\mathbf{Y}\|^2\} < + \infty$, where $E$ denotes the mathematical expectation and $\|\cdot\|$ is the standard Euclidean norm (in $\mathbb{R}^q$). The polynomial chaos expansion (PCE) of $\mathbf{Y}$ is then written as 
% Polynomial Chaos Expansion (PCE) is used in this paper to construct computationally cheaper surrogate models to the assumed ground truth, finite element analysis model. This technique can represent any given black-box model with smooth and bounded outputs as a linear combination of basis functions which are polynomials, to varying levels of accuracy depending on the degree of the polynomial basis functions permitted and the number of samples provided for fitting. The polynomials are crafted to be orthogonal with respect to the joint input probability density and this allows each of the basis polynomials to capture a different independent piece of information in some abstract sense about the relation between the input and output space, maximizing the information extracted from each sample.
% More rigorously, for some given random input vector \( \mathbf{X} \) and associated joint probability density function \( J(\mathbf{X}) \), PCE aims to express the uncertain output \( \mathbf{Y} \) as a function of \( \mathbf{X} \) in the form:
\begin{equation}
    \mathbf{Y} = \sum_{n=0}^{\infty} a_n \Phi_n(\mathbf{X})
\end{equation}
where $\{\Phi_n\}_{n \geq 0}$ are multivariate polynomials that are orthonormal with respect to $P_{\mathbf{X}}$ (that is, $E\{\Phi_n(\mathbf{X})\Phi_{n'}(\mathbf{X})\} = \delta_{nn'}$, where $\delta_{nn'}$ is the Kronecker delta), and $\{a_n\}_{n \geq 0}$ are expansion coefficients \citep{ghanem2003stochastic,ghanem2017handbook}. The above representation defines a surrogate model that, once calibrated, enables the characterization of $\mathbf{Y}$. The coefficients can be computed by exploiting the orthogonality of the Hilbertian basis:
\begin{equation}\label{eq:def-an}
    a_n = E\{f(\mathbf{X}) \Phi_n(\mathbf{X})\} = \int_{\mathbb{R}^m} f(\mathbf{x}) \Phi_n(\mathbf{x}) p_{\mathbf{X}}(\mathbf{x}) d\mathbf{x}
\end{equation}
% \[a_n = \frac{\int f(\mathbf{X}) \Phi_n(\mathbf{X}) J(\mathbf{X}) d\mathbf{X}}{\int \Phi_n^2(\mathbf{X}) J(\mathbf{X}) d\mathbf{X}}\]
% With orthogonal in this context defined as: 
% \[\int_{\Omega} \Phi_i(\mathbf{x}) \Phi_j(\mathbf{x}) J(\mathbf{x}) d\mathbf{x} &= \delta_{ij} \]
% where \( \Omega = \{\mathbf{x} \in \mathbb{R}^m : J(\mathbf{x}) > 0\} \) is the support of the m-dimensional random vector \( \mathbf{X} \) and \( \delta_{ij} \) is the Kronecker delta function.
The choice of the polynomials depends on the distribution on $\mathbf{X}$ \citep{Xiu-2002,Soize-2004}; see \cref{tab:dist_PCE}.
\begin{table}[h]
    \centering
    \begin{tabular}{cc} 
        \toprule
        Distribution & Basis Polynomials \\
        \midrule        
        Gaussian & Hermite \\
        Uniform & Legendre \\
        Gamma & Laguerre \\
        Beta & Jacobi \\
        \bottomrule
    \end{tabular}
    \caption{Some standard distributions and their associated basis polynomials.}
    \label{tab:dist_PCE}
\end{table}
For arbitrary distributions, families of polynomial bases can be constructed via ad hoc orthonormalization techniques; see, e.g., \citep{Perrin2012}. The Stieltjes procedure \citep{stieltjes1884quelques} is implemented in the Python package Chaospy \citep{FEINBERG201546}.

In practice, the PCE representation is truncated by restricting the polynomial order. Adopting a simplified notation (in lieu of the standard notation based on multi-indices), we write 
% depends on the truncation of the infinite series, which in practice, is often truncated at some low order \( N \). This results in the finite sum below, which represents the surrogate models we create and compare in this paper.
%\begin{equation} \mathbf{Y} = f(\mathbf{X}) \approx \sum_{n=0}^{N} a_n \Phi_n(\mathbf{X}) \label{eq1} \end{equation}
\begin{equation}
    \mathbf{Y} \approx \sum_{n=0}^{N} a_n \Phi_n(\mathbf{X})
\end{equation}
The above equation defines is a mean-square convergent approximation, implying that a convergence analysis must be performed with respect to $N$.

In this work, PCE expansions are generated with the Python package Chaospy, which is a general purpose uncertainty quantification toolkit \citep{FEINBERG201546}. Some strategies to compute the PCE coefficients are reviewed in the next section.
%%%%%%%%%%%%%%%%%%%%%%%%%%%%%%%%%%%%%%%%%%%%%%%%%%%%%%%%%%%%%%%%%%%%%%%%%%%%%%%%%%%%%%%%%%%%%%%%%%%%%%%%%%%%%%%%%%%%%
\subsubsection{Evaluation of the Chaos Coefficients}
\label{subsubsec:spectral projection}
There exist various techniques to compute the set of coefficients $\{a_n\}_{n \geq 0}$, including intrusive and non-intrusive techniques; see \citep{le2010spectral,ghanem2017handbook} for reviews. 
% UQ aims to approximate $P(\mathbf{y})$ through sampling and surrogate modeling methods. One of these UQ methods is spectral projection. This technique has broad applications to many domains of engineering and science and is particularly useful when the model $f$ is complex. The benefits of spectral projection as a UQ technique are that (1.) it is non-intrusive, requiring no knowledge of the internal workings of the model $f$, (2.) it can be very efficient in sampling, requiring orders of magnitude less samples compared to simpler methods like Monte Carlo sampling, and (3.) it is compatible with surrogate modeling methods like polynomial chaos expansion, allowing for a construction of substantially cheaper alternatives to computationally expensive and complex (for example finite element analysis) models. 
% spectral projection is a method for strategically selecting points \( \mathbf{X}_i \) in the input space to evaluate the model \( f \) at to maximize, in some abstract sense, the information gained about the model and output space per sample. 
In this paper, we consider the following techniques to compute the chaos coefficients according to Eq.~\eqref{eq:def-an} (and rely on their implementation in the Python package Chaospy).
\begin{itemize}
    \item \textbf{Monte Carlo sampling}: in this case, the mathematical expectation in Eq.~\eqref{eq:def-an} is estimated through 
    \begin{equation}
        a_n \approx \frac{1}{M} \sum_{i = 1}^M f(\mathbf{X}_i) \Phi_n(\mathbf{X}_i)
    \end{equation}
    This strategy presents a low convergence rate (note that this drawback can be partially circumvented using more efficient sampling strategies), which is however independent of $m$ (the dimension of the stochastic input). It remains applicable when the forward model $f$ is reasonably cheap to evaluate. 
    
    % In this sampling strategy, points \( \mathbf{X}_i \) are drawn randomly according to the probability density function \( J(\mathbf{X}) \) but instead of inferring \( P(\mathbf{Y}) \) from the input and output pairs, the pairs are fitted to the PCE surrogate model which can then be sampled relatively cheaply and with high fidelity to the true model \( f \)  to recover an approximate output probability density function \( P_{PCE}(\mathbf{Y}) \).

    \item \textbf{Quadrature rule}:
    Alternatively, the integral in Eq.~\eqref{eq:def-an} can be evaluated using a deterministic (e.g., Gaussian) quadrature rule:
    \begin{equation}
        a_n \approx \sum_{i=1}^{N_Q} w_i f(\mathbf{X}_i)
    \end{equation}
    where $\{w_i\}_{i = 1}^{N_Q}$ and $\{\mathbf{X}_i\}_{i = 1}^{N_Q}$ denote the weights and nodes of the $N_Q$-point cubature. Such rules are typically formulated using a tensorization of one-dimensional cubatures when $m >1$.    
    % In quadrature rule sampling, input points ("nodes") \( \mathbf{X}_i \) are selected at the roots of the orthogonal PCE polynomials \( \Phi_n(\mathbf{X}) \). 
    % These nodes are also assigned a weight \( w_i \) which are also derived from \( \Phi_n(\mathbf{X}) \) (with different formulae for different basis sets due to normalization requirements) and signify to the relative influence of each node on the expected value of \(f(\mathbf{X})\). 
    % \[
    % \int f(\mathbf{X}) J(\mathbf{X}) d\mathbf{X} \approx \sum_{i=1}^{N} w_i f(\mathbf{X}_i)
    % \]
    % These nodes, weights, and their evaluations using the model \(f\) are then used to fit the PCE surrogate model, from which we can extract approximations of the true output space probability density \( P_{PCE}(\mathbf{Y}) \). Because the nodes are polynomial roots and weights and nodes are paired, this implies that the number of quadrature nodes and weights are both a function of the truncation order \( N \) of the PCE polynomial basis. Specifically, the number of nodes/weights generated is \((N+1)^m\), an exponential relationship that is often referred to as the "curse of dimensionality". This \( N \) from here onwards will be referred to as the degree of the quadrature.
    
    \item \textbf{Smolyak sparse grid}:
    In order to circumvent the curse of dimensionality arising in tensor product formula, a sparse grid can be used where a subset of quadrature points is identified based on a given criterion (constraining the sum of all one-dimensional levels of accuracy). This leads to a much smaller number of quadrature points, and enables integration for large values of $m$. Here we use the Smolyak sparse grid introduced in \cite{Smolyak}.
    % The quadrature rule sampling strategy above can be thought of as sampling roots to a full tensor product of 1D polynomials \( \Phi_n(X_1), \Phi_n(X_2) ... \Phi_n(X_m) \) for the m-dimensional input space. This effectively is sampling over a grid of points in m-dimensional space. The sparse grid approach refines this sampling strategy by combining quadrature grid points of different resolutions up to some order N while discarding others to form a subset of the full tensor grid with the number of points scaling as \( N \cdot (\log N)^{m-1} \) which is considerably more impervious to the exponential scaling problem than the quadrature rule sampling strategy.
\end{itemize}

%%%%%%%%%%%%%%%%%%%%%%%%%%%%%%%%%%%%%%%%%%%%%%%%%%%%%%%%%%%%%%%%%%%%%%%%%%%%%%%%%%%%%%%%%%%%%%%%%%%%%%%%%%%%%%%%%%%%%
\subsubsection{Geometry Defects}
\label{subsubsec:m_geo}
As seen in \cref{subsubsec:spectral projection}, the dimensionality of the input space severely affects the amount of computation needed to generate PCE surrogate models. Thus there is intrinsic motivation to keep input space dimensionality low, leading to the natural issue of dealing with increasing resolution and the quadratic scaling in pixel count for 2D metamaterials. It would be infeasible in practice to use a method which requires one dimension input for every pixel in a geometry, and such a method would only work for very crude geometry representations. 

In manufacturing processes, the stochastic nature of defects is not usually a function of the arbitrary resolution one chooses to represent their geometry in, but instead of some scale independent parameter such as lithography laser precision, 3D printing nozzle size, or mill head diameter in CNC machining. Motivated by this, we sought a single or a fixed number of scale invariant parameters which define stochastic defects on the designed geometry, and decided to use a "edge pixel flip proportion" (FP) parameter. Our algorithm for generating geometry defects with this FP parameter is as follows. 
\begin{enumerate}
    \item Specify some design, defect free geometry that is 40 pixels by 40 pixels. 
    \item Identify all the edge pixels in the design geometry, which are hard material pixels that share an edge with a soft material pixel, and vice versa. 
    \item Randomly pick a proportion of edge pixels equal to FP and flip these pixels to the other material.
\end{enumerate}
\begin{figure}[H]
    \centering
    \includegraphics[width=0.75\linewidth]{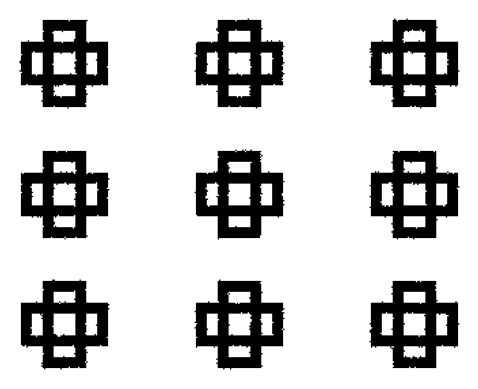}
    \caption{A palette of possible defective shapes for the same FP parameter of 0.05 (5\% flip rate)}
    \label{fig:defect_palette_40}
\end{figure}
The main benefit of this algorithm is that it offers a way to capture many realistic cases of processing defects with a single scale independent parameter. 

There are three potential issues with this algorithm however, the first being that only edge defects are allowed. Although the algorithm could be easily altered to flip a proportion of all pixels, rather than just the edge pixels, we choose to limit the flipping to only edge pixels, motivated by the observation that for many processes, the most serious defects and deviations in manufacturing occur at the edges of geometries. This part of the algorithm can be tweaked without rendering other aspects of the overall methodology obsolete, and so is effectively a non-issue.

Another potential issue is that the output space using FP as an input is not deterministic. For the same FP, multiple defective geometries are possible, leading to the question of whether the single FP parameter is a valid representation of the input geometry. If the possible outputs from a single set of inputs vary wildly, then this would cause catastrophic results in the PCE fitting process, as the PCE process assumes smooth, single valued model functions. However, if the output variation due to different possible defective geometries for the same FP is small relative to the output variation caused by different FP values, then this FP method is a pseudo-deterministic scenario where the variation introduced by multivalued defective geometries can be considered model noise in an otherwise deterministic model and PCE will have a good chance of successfully fitting surrogate models. This variation comparison will be tested in results presented in the \nameref{sec:Results} section. 

A third potential issue is that because only edge pixels are flipped, as resolution increases, less of the shape area is comprised of edge pixels and thus less area is subject to random flipping. This problem was examined closer by looking for convergence in the effects of defects as the shape resolution increased, with results presented in the \nameref{sec:Results} section. 
%%%%%%%%%%%%%%%%%%%%%%%%%%%%%%%%%%%%%%%%%%%%%%%%%%%%%%%%%%%%%%%%%%%%%%%%%%%%%%%%%%%%%%%%%%%%%%%%%%%%%%%%%%%%%%%%%%%%%
\section{Results and Discussion}
\label{sec:Results}
The following sections will detail the performance of the spectral projection methods on 1D and 6D input spaces of material properties, as well as the performance of these uncertainty techniques in a 7D input space scenario including both material parameters and geometry defects of samples. The following sections will also demonstrate the techniques on 4 different types of standard distributions, the uniform, normal, gamma, and beta. For purposes of brevity, not all combinations of distributions and methods are shown in the following results, but the method does generalize to all distributions and additional information and additional figures can be found in the supplementary sections. 
%\label{}
%\subsection{1D Uniform Input Space}
%%%%%%%%%%%%%%%%%%%%%%%%%%%%%%%%%%%%%%%%%%%%%%%%%%%%%%%%%%%%%%%%%%%%%%%%%%%%%%%%%%%%%%%%%%%%%%%%%%%%%%%%%%%%%%%%%%%%%
\subsection{Geometry Defects}
\label{subsec:r_geo}
One of the potential issues with the "flip proportion" (FP) defect generation algorithm was the potential for variation of bandgap outputs due to the interplay between resolution of the geometry and FP. Since our algorithm only flips edge pixels, a higher image resolution implies a smaller proportion of the overall pixels that are considered edge pixels and subject to flipping. To check for the effects of this issue, we hold FP constant and vary the geometry resolution from 10x10 to 100x100 pixels, averaging the outputs of band gap size, location top and bottom over 100 samples due to the pseudo-deterministic nature of generating defects with just one number. Below are the results of this study.
\begin{figure}[H]
	\centering 
	\includegraphics[width=\textwidth]{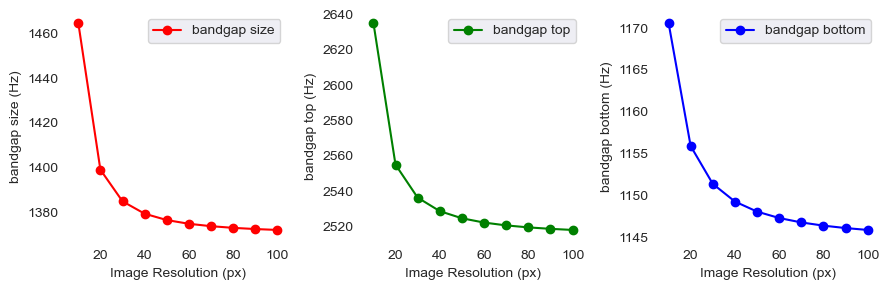}	
	\caption{The average bandgap size, top and bottom locations of 100 defective geometries generated with 0.05 FP at varying image resolutions. Note that the resolution is selected before the flipping of edge pixels.} 
	\label{fig:FEM_convergence_afo_resolution}%
\end{figure}
From figure \ref{fig:FEM_convergence_afo_resolution}, we note a clear convergence of output values as resolution increases, indicating that for low image resolutions of 10 to 30 pixels, we have crude elements which introduces relatively higher error into the FEA model. As the resolution increases to 40 pixels and above, we have such low deviations from each asymptotic output value that it is not computationally worthwhile to increase resolution further. Thus for the rest of the experiments in this paper, geometries were generated and computed with a resolution of 40x40 pixels.

Another potential issue with the FP method was that the pseudo-deterministic nature of the defective geometries, (i.e. the mapping of a FP value to a shape is not one-to-one, but one-to-many), would introduce a variation in outputs that is too large, and confuse the PCE fitting algorithm as the algorithm expects smoothly varying outputs with varying inputs. Thus a study was done to hold the FP constant, and check for the variation introduced by the one-to-many relation at different image resolutions.
\begin{figure}[H]
    \centering
    \begin{subfigure}[b]{0.49\linewidth}
        \centering
        \includegraphics[width=\linewidth]{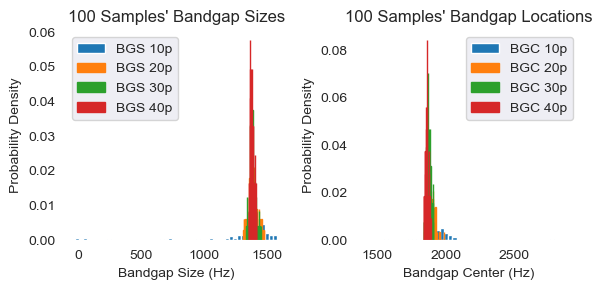}
        \caption{Histogram of computed bandgap size (left) and center location (right) of 100 geometry defect samples generated with the same FP value of 0.05 and paired with the same material properties for resolutions of 10x10 to 40x40 pixels.}
        \label{fig:geo_defect_fp_convergence_10p-40p}
    \end{subfigure}
    \hfill
    \begin{subfigure}[b]{0.49\linewidth}
        \centering
        \includegraphics[width=\linewidth]{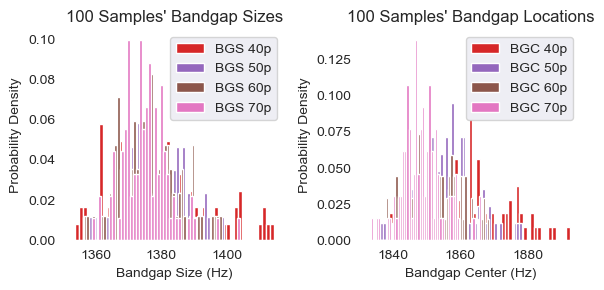}
        \caption{Histogram of computed bandgap size (left) and center location (right) of 100 geometry defect samples generated with the same FP value of 0.05 and paired with the same material properties 40x40 to 70x70 pixels.}
        \label{fig:geo_defect_fp_convergence_40p-70p}
    \end{subfigure}
    \caption{Results of study on variation introduced by non-deterministic FP parameter in generating geometric defects.}
    \label{fig:geo_defect_fp_convergence}    
\end{figure}
The results in \cref{fig:geo_defect_fp_convergence} shows that there is huge variation from different defective geometries generated by the same FP parameter for resolutions of 10 to 30 pixels but low variation at 40 pixels and above. For reference, it was found through simulations that realistic stochastic material property ranges resulted in output ranges for bandgap size and center position that was several hundred Hz. We would want any noise effects like the differences between geometries of the same FP to be well below this magnitude invariation. At under 30 pixels, we have variations in the hundreds to even thousands of Hz, due to each flipped pixel removing or adding a relatively large portion of the overall structure. At over 40 pixel image resolution, we see that like with the previous study, there is convergence asymptotically to some average value for the bandgap size and location. Additionally, there is also convergence asymptotically in output distribution and that distribution has a range on the order of tens of Hertz. This means that the variation for one FP value (noise) is at least an order of magnitude lower relative to the variation introduced by the FP value itself and other input dimensions (signal), and that we can consider FP to be effectively deterministic and use it as a valid representation of geometric defects. Thus for following experiments, each set of sampled material properties will be paired with one defective geometry randomly generated with a sampled FP parameter.
%%%%%%%%%%%%%%%%%%%%%%%%%%%%%%%%%%%%%%%%%%%%%%%%%%%%%%%%%%%%%%%%%%%%%%%%%%%%%%%%%%%%%%%%%%%%%%%%%%%%%%%%%%%%%%%%%%%%%
\subsection{6D Gamma + 1D Beta (Material Properties + Geometry) Input Study}
\label{subsec:7D Gamma Input Space}

In this study, stochastic input parameters include six material properties, assumed to be gamma distributed, and the geometry defect parameter FP. The latter is assumed to follow a beta distribution. The definition of the material property distributions follows the previous works \citep{Guilleminot_Soize_2013,Staber2017} which showed, using information theory, that the gamma distribution constitutes an objective choice in stochastic isotropic elasticity. The choice of beta distribution ensures that the geometry flip proportion parameter takes values between 0 and 1. The distribution parameters are provided in \cref{7D_Gamma_Input_Table}, and the associated histograms are shown in \cref{7d_gamma_input_mc_10000} below: 

\begin{table}[H]
\centering
%\small
\tiny
\caption{7D Gamma Material Property \& Beta Geometry Input Distributions}
\label{7D_Gamma_Input_Table}
\begin{tabularx}{\linewidth}{Xccccc} 
\toprule
Material Property & Distribution & Mean ($\mu$) & Standard Deviation ($\sigma$) & Shape ($\alpha$) & Rate/Shape ($\beta$) \\
\midrule
\( \text{Soft Bulk Modulus } K_{soft}\) & \( \text{Gamma} \) & \( 278 \text{ MPa} \) & \( 0.08\mu \) & \( 1.56\cdot10^2 \) & \( 1.78\cdot10^6 \) \\
\( \text{Hard Bulk Modulus } K_{hard}\) & \( \text{Gamma} \) & \( 152 \text{ GPa} \) & \( 0.02\mu \) & \( 2.50\cdot10^3 \) & \( 6.06\cdot10^7 \) \\
\( \text{Soft Shear Modulus } G_{soft}\) & \( \text{Gamma} \) & \( 72.5 \text{ MPa} \) & \( 0.08\mu \) & \( 1.56\cdot10^2 \) & \( 4.64\cdot10^5 \) \\
\( \text{Hard Shear Modulus } G_{hard}\) & \( \text{Gamma} \) & \( 78.1 \text{ GPa} \) & \( 0.02\mu \) & \( 2.50\cdot10^3 \) & \( 3.13\cdot10^7 \) \\
\( \text{Soft Density } \rho_{soft}\) & \( \text{Gamma} \) & \( 1000 \text{ g/cm$^3$} \) & \( 0.08\mu \) & \( 1.56\cdot10^2 \) & \( 6.4 \) \\
\( \text{Hard Density } \rho_{hard}\) & \( \text{Gamma} \) & \( 8000 \text{ g/cm$^3$} \) & \( 0.02\mu \) & \( 2.50\cdot10^3 \) & \( 3.2 \) \\
\( \text{Geometry Flip Proportion} \) & \( \text{Beta} \) & \( 0.025 \) & \( 0.08\mu \) & \( 1.52\cdot10^2 \) & \( 5.94\cdot10^3 \) \\
\bottomrule
\end{tabularx}
\end{table}

\begin{figure}[H]
	\centering 
	\includegraphics[width=\textwidth]{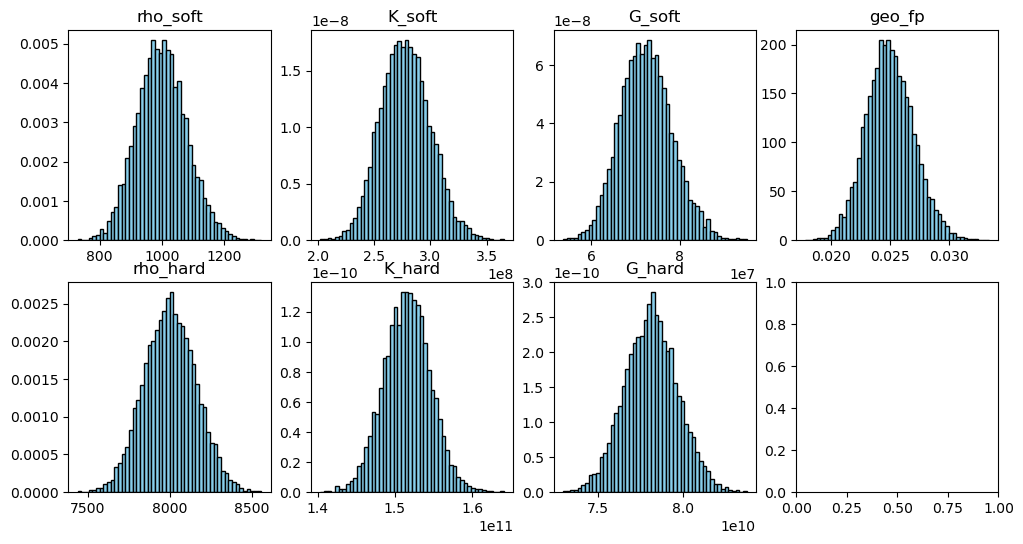}	
	\caption{10000 Monte Carlo samples of the 7D Gamma and Beta input space, for visualization of the input space distribution shapes.} 
	\label{7d_gamma_input_mc_10000}%
\end{figure}
\begin{figure}[H]
	\centering 
	\includegraphics[width=\textwidth]{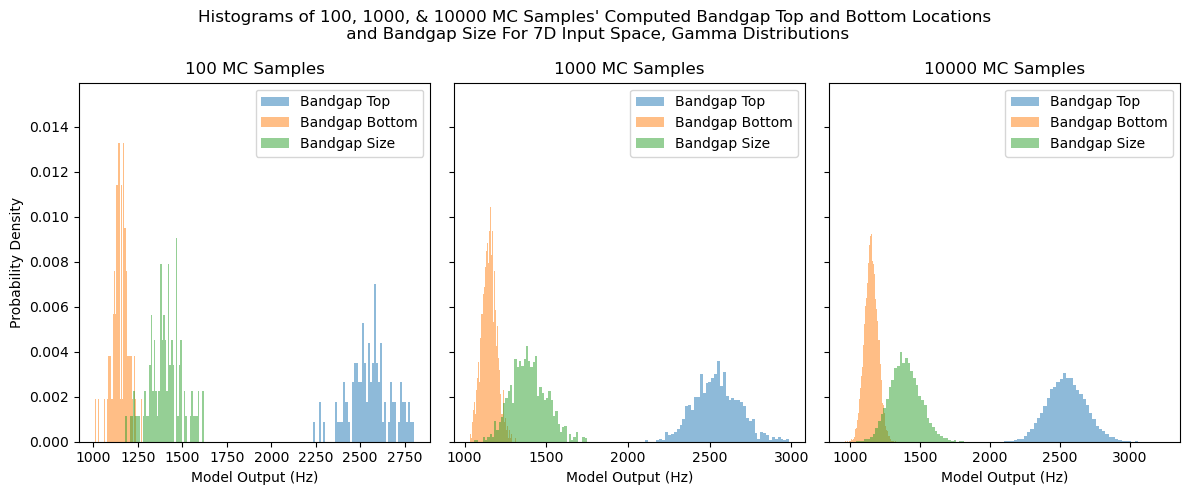}	
	\caption{Histograms of the output variables for the Monte Carlo input datasets, band gap size and location, with location expressed in terms of bandgap top and bottom. Note that these two are combined later into the quantity bandgap center.} 
	\label{bgtbs_trio_hist_7d_gamma}%
\end{figure}
\begin{figure}[H]
	\centering 
        \begin{subfigure}[b]{\textwidth}
            \centering 
            \includegraphics[width=\textwidth]{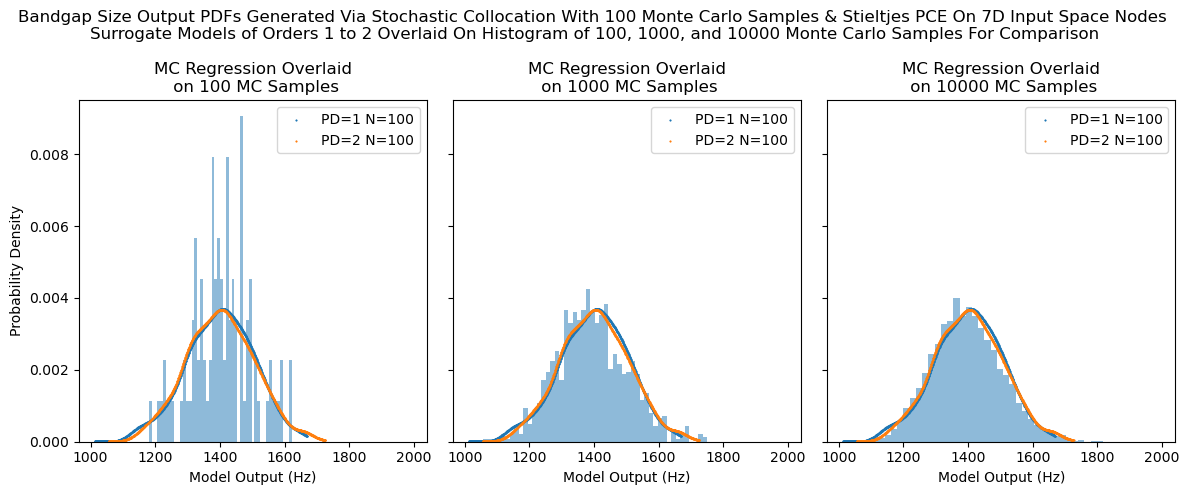}	    
        \end{subfigure}
        \begin{subfigure}[b]{\textwidth}
            \centering
            \includegraphics[width=\textwidth]{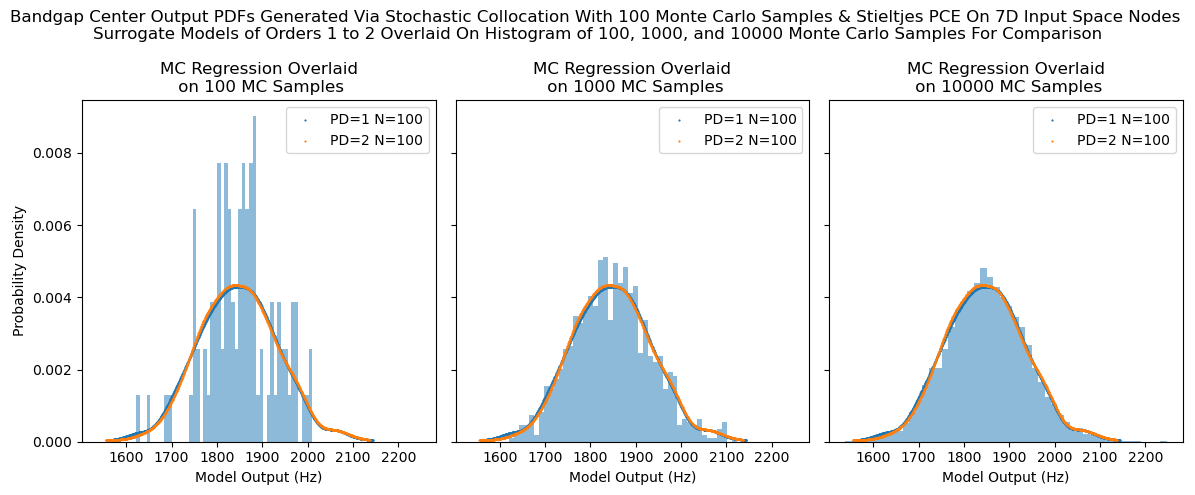}	    
        \end{subfigure}
	\caption{Probability density functions of bandgap size and center from 1st and 2nd degree surrogate models constructed from 100 MC samples, overlaid on histograms of 100, 1000, and 10000 computed MC samples. The curves in each pane, which are generated with KDE on surrogate samples, are the same and it is only the background histograms that change. This indicates that with the same 100 samples as on the left plot panes, the PCE process was able to reconstruct a PDF that matches nearly perfectly to the 10000 sample ground truth.} 
	\label{bgs_trio_hist_quad_fit_7d_gamma}%
\end{figure}
\begin{figure}[H]
	\centering
	\includegraphics[width=\textwidth]{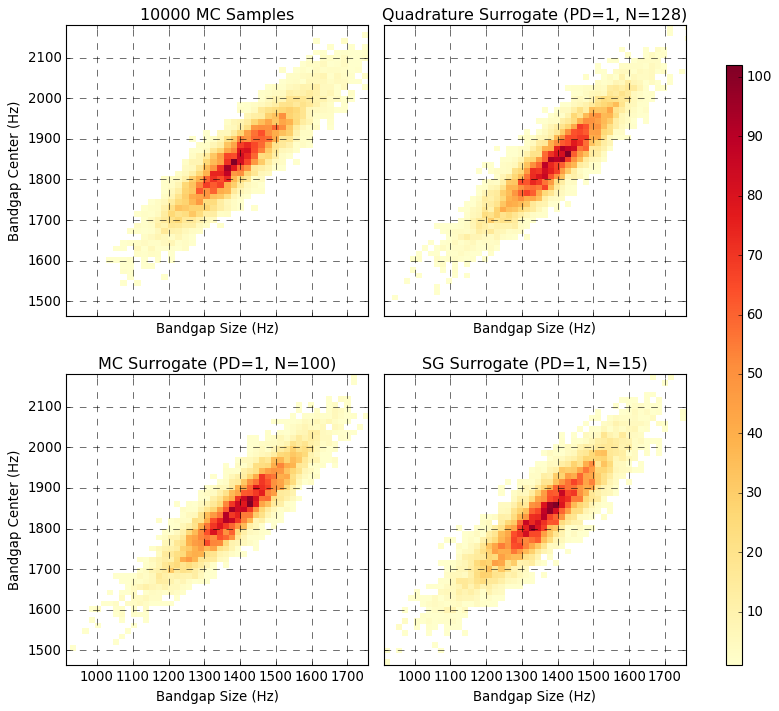}	
        \caption{2D histograms of bandgap sizes and center locations. Top left represents ground truth and is constructed by 10000 Monte Carlo samples. Top right is 10000 samples drawn from the surrogate model constructed from quadrature rule sampling (128 samples). Bottom Left is 10000 samples from the surrogate model constructed from PCE fit on 100 Monte Carlo samples, and bottom right is 10000 samples from the surrogate model constructed from PCE fit on 15 sparse grid samples.}
	\label{fig:2d_hist_7d_input_pd1_quad_gamma}%
\end{figure}

The sampled geometry FP parameters are then used to randomly generate defective geometries to pair with each set of sampled material properties. The typical process and result of the geometry defect generation process is shown in \cref{defect_trio_1st_geo} below.
\begin{figure}[H]
	\centering 
	\includegraphics[width=\textwidth]{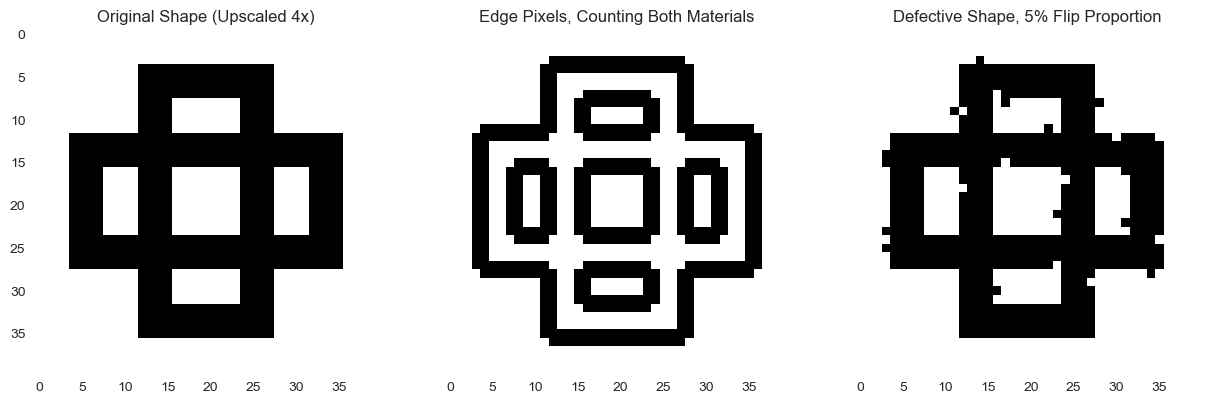}	
	\caption{Left: The defect free designed geometry, scaled up to 40px by 40px. Center: edge pixels of the design geometry, which will be subjected to random flipping at preset proportion of the FP parameter. Right: one resulting defective geometry after flipping edge pixels. Note that this is a representative defect and that different randomly generated defects are (likely) used in each set of sample inputs.} 
	\label{defect_trio_1st_geo}%
\end{figure}
For the three spectral projection sampling strategies, the method parameters and corresponding number of sample points are shown in \cref{table:7D_gamma_sampling_strategies} below. Note that the 10000 Monte Carlo sample set is taken to represent the ground truth for computing the true probability density function (PDF) of the output space (bandgap size and center location). The degree for the quadrature rule approach was set to be 2, since due to the exponential nature of the full tensor grid product, degree 3 or higher would require more points and computation than the 10000 MC samples and would thus be useless as a way to approximate the output space PDF. For the sparse grid approach, we look to maximize computational savings and so choose the lowest grid order as a comparison point.
\begin{table}[H]
    \centering
    \small
    \caption{Polynomial degree and sample points for each of the spectral projection methods}
    \label{table:7D_gamma_sampling_strategies}
    \begin{tabular}{ccc} 
        \toprule
        Sampling Method & Degree & Number of Points \\
        \midrule
        Monte Carlo & N/A & 100, 1000, 10000 \\
        Quadrature Rule & 1, 2 & 128, 2187 \\
        Sparse Grid & 1 & 15 \\
        \bottomrule
    \end{tabular}
\end{table}
For visualization purposes, some of the results of running the FEA model on the above datasets are shown below in \cref{bgtbs_trio_hist_7d_gamma}. It would be difficult for the human eye to perceive the exact shape of the output distributions until some number of samples between $10^3$ or $10^4$, and so the purpose of this study is to see if the same output distribution shape can be captured for much fewer than $\sim 10^3$ samples.
These outputs and inputs are then fed into PCE models for fitting. Only one set of representative fit results is shown in \cref{bgs_trio_hist_quad_fit_7d_gamma} for visualization purposes and brevity. The fit process for all the datasets in \cref{table:7D_gamma_sampling_strategies} is the same and generates comparable results. The PCE surrogate fit however, will fail in cases where there are insufficient samples to fit all the polynomial coefficients (underdetermined problem), but usually does not suffer from overfitting issues as the PCE model does not aim to converge pointwise to the true model, instead only converging in PDF.
%\todo{remove titles from figures, consolidate captions, change x-axis to differentiate top row from bottom.}
% \begin{figure}[H]
%     \centering
%     \begin{subfigure}[b]{0.85\textwidth}
%         \centering
%         \includegraphics[width=\linewidth]{Figures/bgs_trio_hist_quad_fit_7d_gamma.png}
%         %\caption{First subfigure}
%         %\label{fig:sub1}
%     \end{subfigure}    
%     \begin{subfigure}[b]{0.85\textwidth}
%         \centering
%         \includegraphics[width=\linewidth]{Figures/bgs_trio_hist_mc_fit_7d_n100_gamma.png}
%         %\caption{Second subfigure}
%         %\label{fig:sub2}
%     \end{subfigure}    
%     \begin{subfigure}[b]{0.85\textwidth}
%         \centering
%         \includegraphics[width=\linewidth]{Figures/bgs_trio_hist_sg_fit_7d_n15_gamma.png}
%         %\caption{Third subfigure}
%         %\label{fig:sub3}
%     \end{subfigure}       
%     \caption{Main caption for all subfigures}
%     \label{fig:main}
% \end{figure}
In \cref{fig:2d_hist_7d_input_pd1_quad_gamma} below we compare the fit results of the three spectral projection sampling strategies and PCE surrogates with 10000 Monte Carlo computed samples which represents ground truth. One can see that with orders of magnitude fewer samples, we have very closely matched the probability distribution of the actual output space. Even the sparse grid samples numbering only 15 points, was able to faithfully represent, albeit with slight differences, the probability distribution of the output space.
%%%%%%%%%%%%%%%%%%%%%%%%%%%%%%%%%%%%%%%%%%%%%%%%%%%%%%%%%%%%%%%%%%%%%%%%%%%%%%%%%%%%%%%%%%%%%%%%%%%%%%%%%%%%%%%%%%%
\subsection{7D Gaussian Input Space Study}
\label{subsec:7D Gaussian Input Space}
This study demonstrates the applicability of the spectral projection and PCE methods to a different set of distributions, with different material properties, and a different unit cell geometry. Like the previous study, the input space consists of six stochastic material properties and a geometry defect parameter. However unlike previously, all 7 inputs now have truncated normal distributions bounded by four standard deviations above and below the mean. This distribution choice represents another popular choice in the literature for representing material parameters and are detailed in \cref{7D_Gaussian_Input_Table} and \cref{7d_input_mc_10000_2nd_geo} below. The material properties have also been swapped in this study from Bulk and Shear moduli to their counterparts, the Young's Modulus and Poisson ratio. While the Young's Modulus numerically is on the same scale as the Bulk and Shear moduli (GPa and MPa for the hard and soft materials respectively), the Poisson ratio is on a completely different scale, bounded between 0 and 0.5, and represents a natural opportunity to test to see if the method and software packages can handle distributions at substantially different scales, which in theory should be of no issue.

\begin{table}[H]
\centering
%\small
\tiny
\caption{7D Gaussian Input Space Material \& Geometry Property Distributions and Parameters}
\label{7D_Gaussian_Input_Table}
\begin{tabularx}{\linewidth}{Xccccc} 
\toprule
Material Property & Distribution & Mean ($\mu$) & Standard Deviation ($\sigma$) & Lower Trunc. & Upper Trunc. \\
\midrule
\( \text{Soft Stiffness} \) & \( \text{Truncated Normal} \) & \( 200 \text{ MPa} \) & \( 0.08\mu \) & \( -4\sigma \) & \( 4\sigma \) \\
\( \text{Hard Stiffness} \) & \( \text{Truncated Normal} \) & \( 200 \text{ GPa} \) & \( 0.02\mu \) & \( -4\sigma \) & \( 4\sigma \) \\
\( \text{Soft Density} \) & \( \text{Truncated Normal} \) & \( 1000 \text{ g/cm$^3$} \) & \( 0.08\mu \) & \( -4\sigma \) & \( 4\sigma \) \\
\( \text{Hard Density} \) & \( \text{Truncated Normal} \) & \( 8000 \text{ g/cm$^3$} \) & \( 0.02\mu \) & \( -4\sigma \) & \( 4\sigma \) \\
\( \text{Soft Poisson Ratio} \) & \( \text{Truncated Normal} \) & \( 0.38 \) & \( 0.02\mu \) & \( -4\sigma \) & \( 4\sigma \) \\
\( \text{Hard Poisson Ratio} \) & \( \text{Truncated Normal} \) & \( 0.28 \) & \( 0.02\mu \) & \( -4\sigma \) & \( 4\sigma \) \\
\( \text{Geometry Flip Proportion} \) & \( \text{Truncated Normal} \) & \( 0.025 \) & \( 0.08\mu \) & \( -4\sigma \) & \( 4\sigma \) \\
\bottomrule
\end{tabularx}
\end{table}

\begin{figure}[H]
	\centering 
	\includegraphics[width=\textwidth]{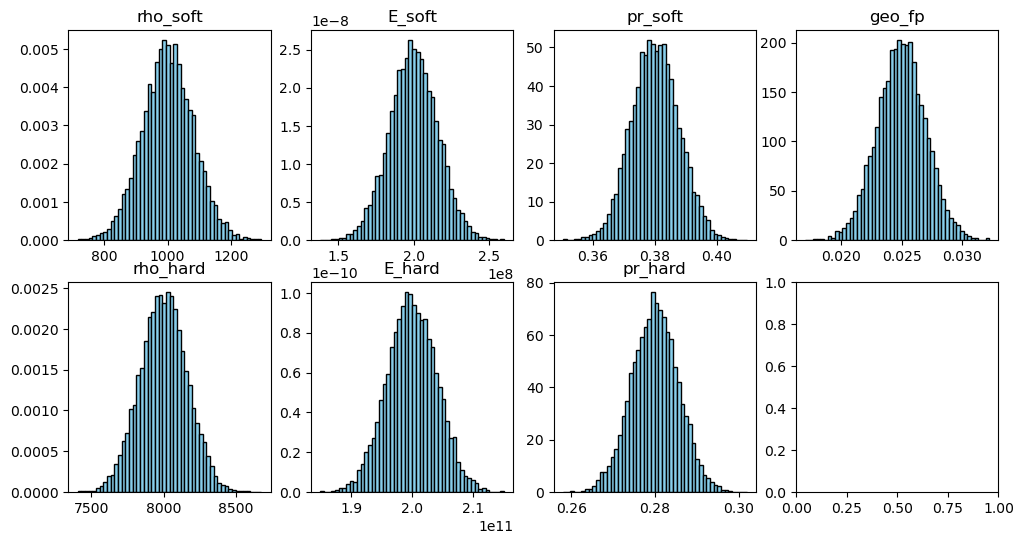}	
	\caption{10000 Monte Carlo samples of the 7D Gaussian input space, for visualization of the input space distribution shapes.} 
	\label{7d_input_mc_10000_2nd_geo}%
\end{figure}
\begin{figure}[H]
	\centering 
	\includegraphics[width=\textwidth]{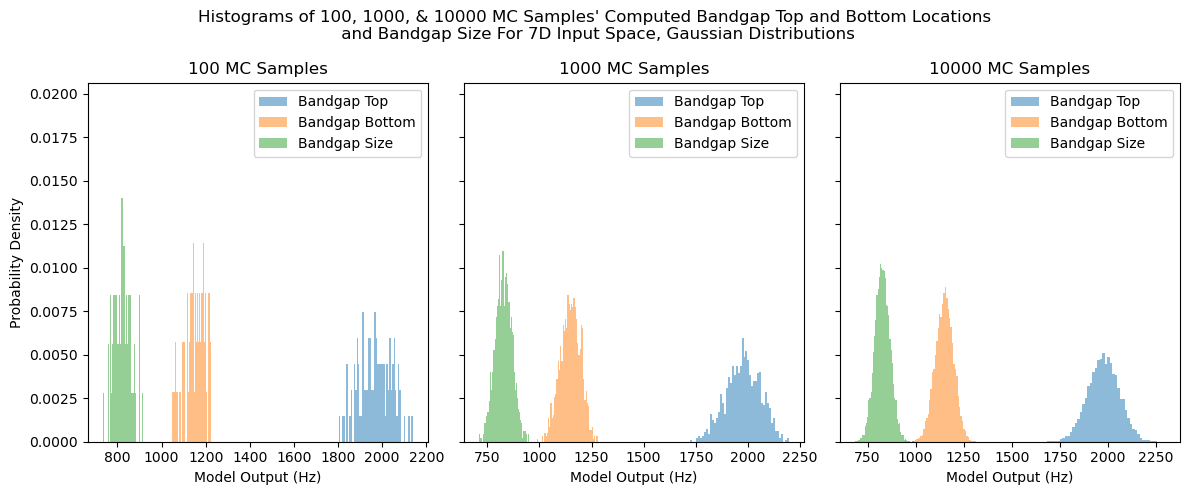}	
	\caption{Histograms of the output variables for the Monte Carlo input datasets, band gap size and location, with location expressed in terms of bandgap top and bottom. Note that these two are combined later into the quantity bandgap center.} 
	\label{bgtbs_trio_hist_7d_gaussian_2nd_geo}%
\end{figure}
\begin{figure}[H]
	\centering 
        \begin{subfigure}[b]{\textwidth}
            \centering 
            \includegraphics[width=\textwidth]{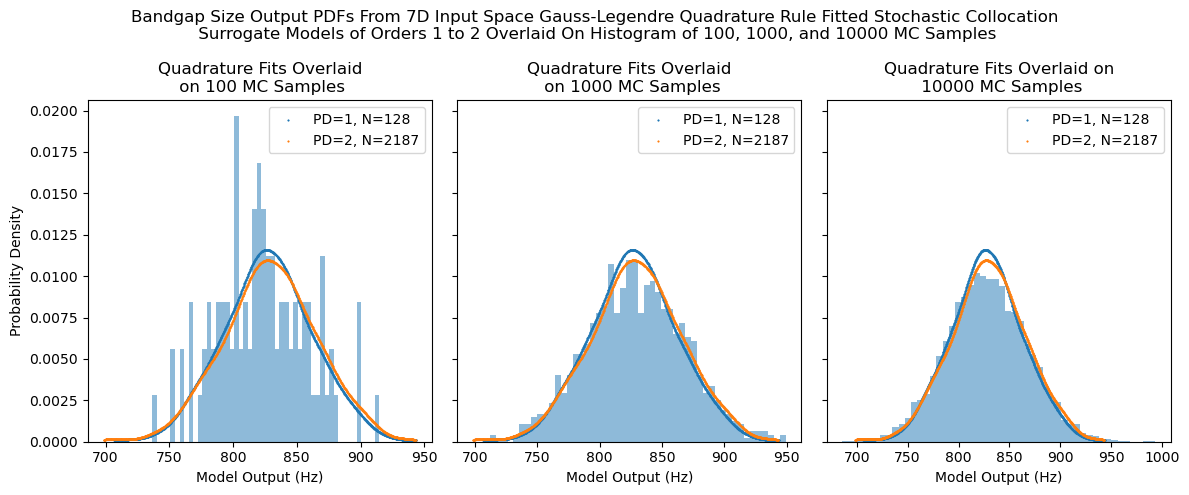}	    
        \end{subfigure}
        \begin{subfigure}[b]{\textwidth}
            \centering
            \includegraphics[width=\textwidth]{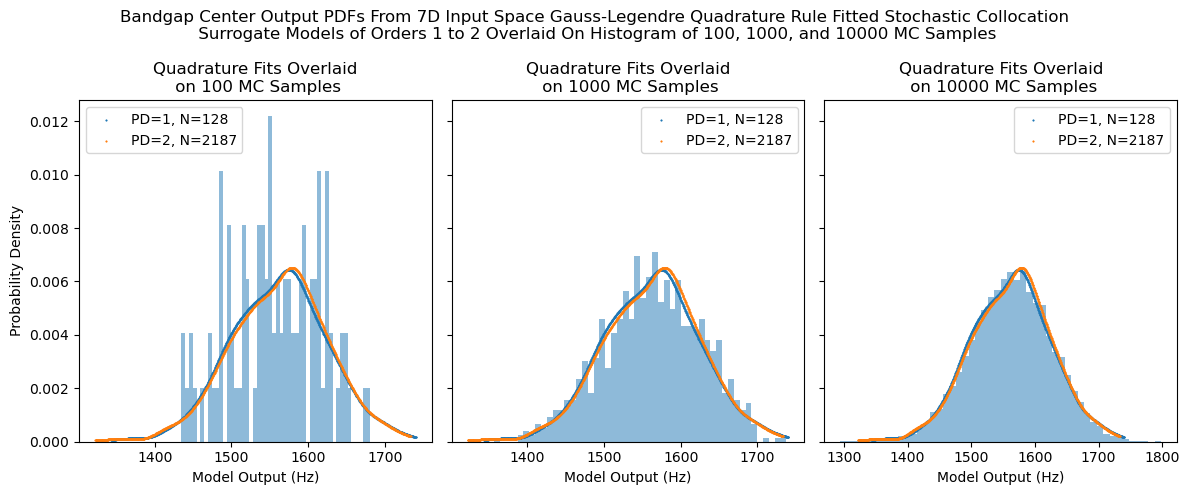}	    
        \end{subfigure}
	\caption{Probability density functions of bandgap size and center from 1st and 2nd degree surrogate models constructed from 1st and 2nd order quadrature rule samples (128 \& 2187 samples respectively), overlaid on histograms of 100, 1000, and 10000 computed MC samples. Note that the curves in each pane, which are generated with KDE on surrogate samples, are the same and it is only the background histograms that change. We can infer that the 2nd order quadrature rule sampling is not necessary as it has a very similar PDF to the 1st order surrogate, which for 128 samples, appears to match the true PDF very closely.} 
	\label{bgs_trio_hist_quad_fit_7d_2nd_geo}%
\end{figure}

Like before, the sampled geometry FP parameters are used to randomly generate defective geometries to pair with each set of sampled material properties. The typical process and result of the geometry defect generation process is shown in \cref{defect_trio_2nd_geo} below.
\begin{figure}[H]
	\centering 
	\includegraphics[width=\textwidth]{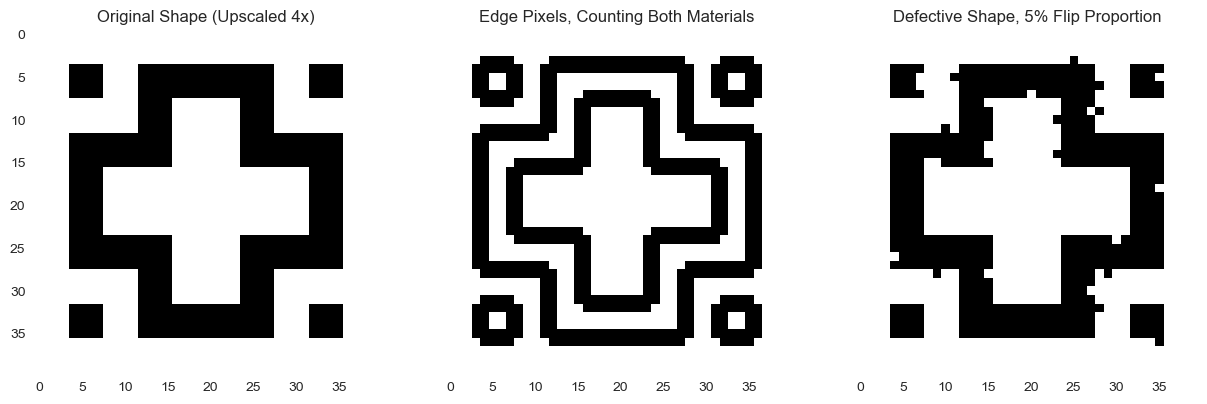}	
	\caption{Left: The defect free designed geometry, scaled up to 40px by 40px. Center: edge pixels of the design geometry, which will be subjected to random flipping at preset proportion of the FP parameter. Right: one resulting defective geometry after flipping the preset proportion of the edge pixels. Note that this is a representative defect and that different randomly generated defects are (likely) used in each set of sample inputs.} 
	\label{defect_trio_2nd_geo}%
\end{figure}
The three spectral projection sampling strategies, shown in \cref{table:7D_gauss_sampling_strategies} below are the same as in the previous study. Like before, the 10000 Monte Carlo sample set is taken to represent the ground truth for computing the true probability density function (PDF) of the output space (bandgap size and center location). 
\begin{table}[H]
    \centering
    \small
    \caption{Polynomial degree and sample points for each of the spectral projection methods}
    \label{table:7D_gauss_sampling_strategies}
    \begin{tabular}{ccc} 
        \toprule
        Sampling Method & Degree & Number of Points \\
        \midrule
        Monte Carlo & N/A & 100, 1000, 10000 \\
        Quadrature Rule & 1, 2 & 128, 2187 \\
        Sparse Grid & 1 & 15 \\
        \bottomrule
    \end{tabular}
\end{table}
For visualization purposes, some of the results of running the FEA model on the above datasets are shown below in \cref{bgtbs_trio_hist_7d_gaussian_2nd_geo}. Like previously, it is difficult for the human eye to perceive the exact shape of the output distributions until sample number reaches somewhere near $10^3$ or $10^4$, and so we try again to see if the true output distribution shape can be captured for much fewer than $\sim 10^3$ samples.
These outputs and inputs are fed into PCE models for fitting, which like previously, produced very similar results for each dataset, so for the purposes of visualization and brevity, only one set of representative surrogate fit results are shown in \cref{bgs_trio_hist_quad_fit_7d_2nd_geo}.

\begin{figure}[H]
	\centering
	\includegraphics[width=\textwidth]{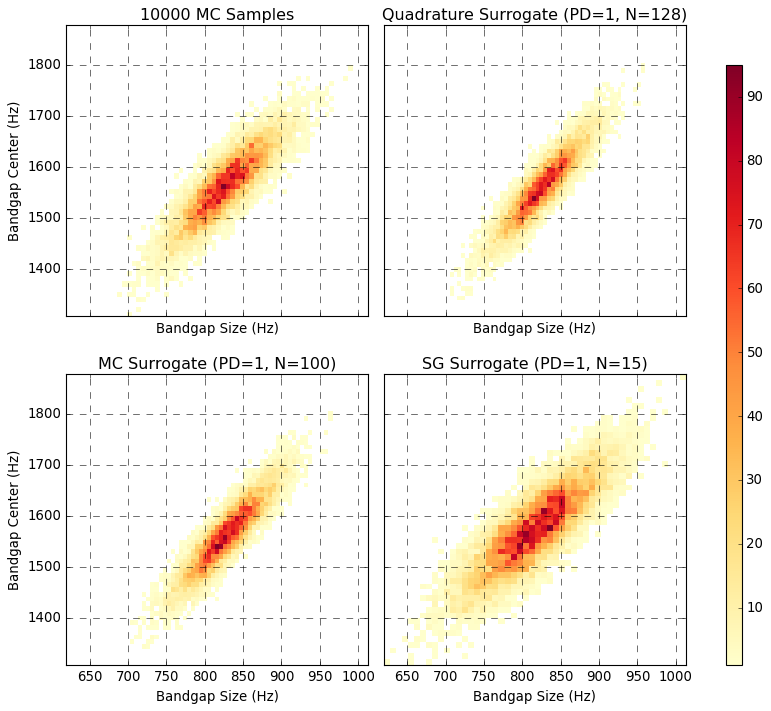}	
        \caption{2D histograms of bandgap sizes and center locations. Top left represents ground truth and is constructed by 10000 Monte Carlo samples. Top right is 10000 samples drawn from the surrogate model constructed from quadrature rule sampling (128 samples). Bottom Left is 10000 samples from the surrogate model constructed from PCE fit on 100 Monte Carlo samples, and bottom right is 10000 samples from the surrogate model constructed from PCE fit on 15 sparse grid samples.}
	\label{fig:2d_hist_7d_input_pd1_quad_2nd_geo}%
\end{figure}

In \cref{fig:2d_hist_7d_input_pd1_quad_2nd_geo} below we compare the fit results of the three spectral projection sampling strategies and their PCE surrogates with 10000 Monte Carlo computed samples which represents ground truth. Like with the previous study, we have very closely matched the probability distribution of the actual output space with order(s) of magnitude fewer samples. The sparse grid surrogate, using only 15 points, deviates further from the true distribution than the previous study, but is still reasonably good at about 12.5\% wider spread in the domain on both outputs.

%%%%%%%%%%%%%%%%%%%%%%%%%%%%%%%%%%%%%%%%%%%%%%%%%%%%%%%%%%%%%%%%%%%%%%%%%%%%%%%%%%%%%%%%%%%%%%%%%%%%%%%%%%%%%%%%%%%
\subsection{1D Uniform}
\label{subsec:1D Uniform Input Space Study}
\begin{figure}[H]
	\centering
	\includegraphics[width=\textwidth]{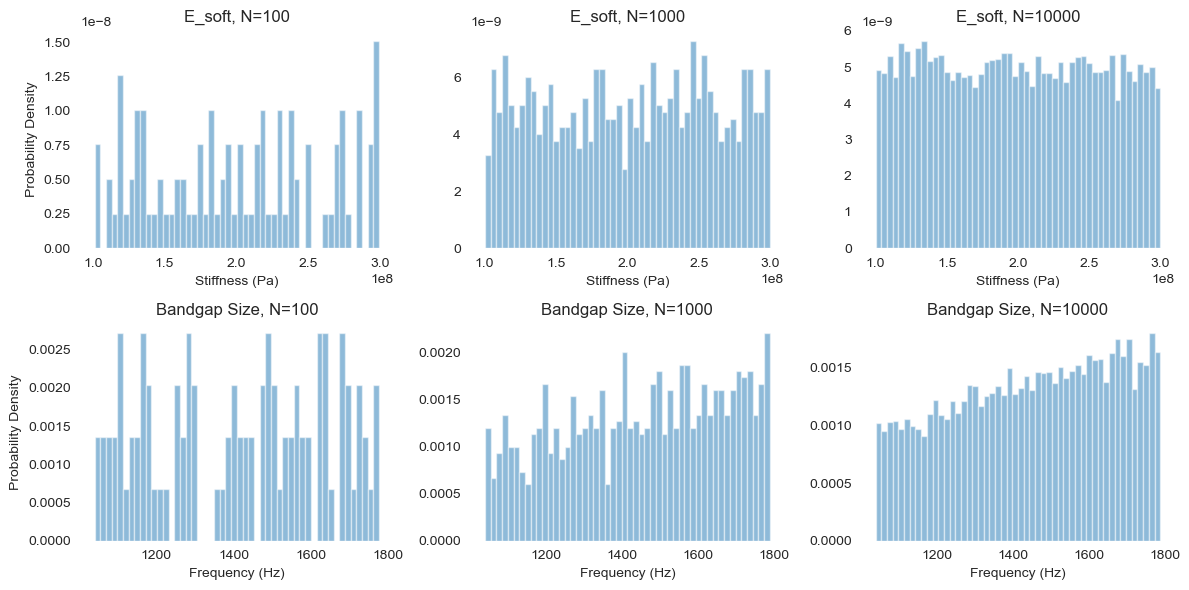}	
        \caption{Histograms of 100, 1000, and 10000 randomly sampled soft material stiffness (top row) and corresponding computed bandgap sizes (bottom row).}
	\label{fig:1d_input_E_soft_q_bgs_trio_hist}%
\end{figure}

In this study, we examine how the Monte Carlo and quadrature rule sampling strategies perform in the low dimension case, with just one input. Note that the sparse grid strategy does not really make sense to be employed here as it essentially degenerates into the same strategy as quadrature rule for the 1D case. The distribution chosen for this study is the uniform, which gives us yet another comparison distribution for the methods' performance, but also represents scenarios where one is able to produce or choose a material with a tunable property, and wants to ascertain the effects of all setpoints of that property on an output of interest. In \cref{fig:1d_input_E_soft_q_bgs_trio_hist} below we see the inputs and FEA model outputs of three datasets. Because the process works the same way and achieves comparable levels of performance regardless of which material property is varied, for the sake of brevity and visualization, we present only the case of varying the soft material stiffness \(E_{soft}\).
For the quadrature rule sampling strategy, we will look at orders \(N = 2 \text{ to } 5\), which corresponds in the 1D case (\(m=1\)) to a number of points \(n = (N+1)^m = 3 \text{ to } 6\).
\begin{figure}[H]
	\centering
	\includegraphics[width=\textwidth]{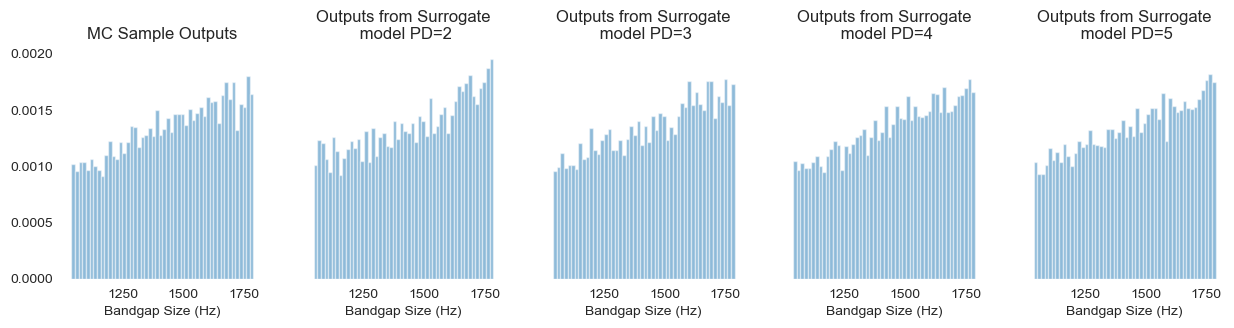}	
        \caption{Histograms of 10000 output samples drawn from FEA on Monte Carlo samples, and PCE surrogate models fitted to 3,4,5, and 6 quadrature rule points respectively.}
	\label{fig:1d_input_E_soft_q_bgs_quinto_hist}%
\end{figure}
From the results above, we can see that at orders 2 and above, the surrogate model output distribution is virtually indistinguishable from the true output probability density. This is a remarkable result as it indicates that with just 3 to 6 samples (and appropriate weights), we are able to capture the underlying probability distribution governing these samples. 

%%%%%%%%%%%%%%%%%%%%%%%%%%%%%%%%%%%%%%%%%%%%%%%%%%%%%%%%%%%%%%%%%%%%%%%%%%%%%%%%%%%%%%%%%%%%%%%%%%%%%%%%%%%%%%%%%%%
\section{Conclusions}
%\todo{shorten by moving some material to results, usually does not refer back to other sections (standalone)}
%\todo{Some lines about importance of work and a line above future work}

%%%%%%%%%%%%%%%%%%%%%%%%%%%%%%%%%%%%%%%%%%%%%%%%%%%%%%%%%%%%%%%%%%%%%%%%%%%%%%%%%%%%%%%%%%%%%%%%%%%%%%%%%%%%%%%%%%%
\subsection{spectral projection \& PCE}
From the studies in \cref{subsec:7D Gamma Input Space} and \cref{subsec:7D Gaussian Input Space}, 
we found that all of the three spectral projection sampling strategies: MC sampling, quadrature rule, and sparse grid, were able to capture the probability distribution of the 2D output space of bandgap size and location given a 7D input space of 6 material properties and a geometry defect parameter. We see the most extreme sample size savings for the sparse grid approach, which with \(\sim 15\) points, was able to capture fairly accurately the output probability distribution. The sparse grid approach however, due to its sparsity, may not work as well for output landscapes with many local features, which may be missed by the sparse grid points. Quadrature rule sampling, which returns a full grid instead of a sparse grid of points in the input space, is less likely to miss local features, but scales exponentially with input dimension and so is not practical for much higher of a input space dimensionality to what is done in this paper's studies. Monte Carlo sampling paired with PCE is relatively less effective at lower dimensions, but because it does not scale directly with the input space dimensionality (it can scale through indirect means, such as model complexity which typically increases with input dimensionality), it performs better at higher dimensions, slightly beating out quadrature rule in our studies at 100 points to achieve good fit over 128 points for the 1st degree quadrature rule. As seen in \cref{subsec:1D Uniform Input Space Study} however, in the 1D case, quadrature rule spectral projection dominates, requiring only a handful of samples (3-6) to faithfully capture the output probability distribution of a FEA model. These results are in line with prevailing wisdom about the different spectral projection sampling strategies. In general, spectral projection and PCE is a powerful tool for analysing acoustic metamaterial performance characteristics in the context of stochastic material properties or geometry.
%%%%%%%%%%%%%%%%%%%%%%%%%%%%%%%%%%%%%%%%%%%%%%%%%%%%%%%%%%%%%%%%%%%%%%%%%%%%%%%%%%%%%%%%%%%%%%%%%%%%%%%%%%%%%%%%%%%
\subsection{Geometry Encoding Schema}
Our approach to encoding geometry defects may be of utility to those who wish to incorporate geometry stochasticity into their uncertainty quantification analysis. In general, there are three types of approaches, which suffer from different advantages and disadvantages. The brute force way of representing each pixel of a given geometry is the most straight forward, offers the most control, but is computationally very costly, with input dimensionality scaling quadratically with resolution (typically given as a length of pixels), which when coupled with the exponential scaling of some sampling strategies with input dimensionality, renders only Monte Carlo sampling potentially an option for this approach, if one can tolerate the innate quadratic scaling. A second class of approaches is to encode the geometry as a set of latent features using some transformer, an approach that is popular in many machine learning endeavors and detailed in \citep{ChenWei2022}. This approach can be highly efficient, and theoretically can be the most efficient with appropriate regularization in the loss function penalizing redundant or unnecessary latent features. However, the main drawback of this approach is that often the number of latent features and their physical meaning is not interpretable. It can also be extremely difficult for technicians and engineers to examine samples in a realistic manufacturing scenario and ascribe probability distributions to each of the latent features, a step that is necessary to leverage the power of PCE and spectral projection. The final class of approaches are those like the algorithm detailed in methodology \cref{subsubsec:m_geo}. These are interpretable simplifications, that rely on some symmetry or intuition of the physical world in order to reduce the set from all possible deformations into the set of those likely or interesting to a given problem. For our problem setup, it was the intuition of how such acoustic metamaterials would be manufactured that informed restricting geometric defects to occur at the edges. This resulted in a viable method which was pseudo-deterministic (variations for the same flip proportion << variations for different flip proportion parameters and for different material properties.), and effective at constraining the curse of dimensionality problem with our UQ methods, removing any scaling relationship between input space dimension and geometry image resolution.
%%\label{}

\section*{Acknowledgements}
Special thanks to Alexander C. Ogren of the California Institute of Technology for providing the finite element analysis code which served as the true model to compute samples from and compare with.

%% The Appendices part is started with the command \appendix;
%% appendix sections are then done as normal sections
\appendix
\section{Supplementary Figures}
\subsection{7D Gamma \& Beta Distribution Experiment}

\subsection{7D Gaussian Distribution Experiment}
%% \label{}

%\section{Appendix title 2}
%% \label{}

%% If you have bibdatabase file and want bibtex to generate the
%% bibitems, please use
%%
\bibliographystyle{elsarticle-harv}
\bibliography{references}
%\todo{Add some references for metamaterials, or UQ}
%% else use the following coding to input the bibitems directly in the
%% TeX file.

%%\begin{thebibliography}{00}

%% \bibitem[Author(year)]{label}
%% For example:

%% \bibitem[Aladro et al.(2015)]{Aladro15} Aladro, R., Martín, S., Riquelme, D., et al. 2015, \aas, 579, A101

%%\end{thebibliography}

\end{document}